\documentclass[11pt,a4paper]{article}
\pdfoutput=1

\usepackage{jheppub}

\usepackage{mciteplus}

\usepackage[T1]{fontenc}
\usepackage[utf8]{inputenc}
\usepackage{amsmath}
\usepackage{amssymb}
\usepackage{hyperref}
\usepackage{graphicx} 

\arxivnumber{1211.5370}

\title{UV completions of flavour models and large $\theta_{13}$}
\author[a]{Ivo de Medeiros Varzielas}
\emailAdd{ivo.de@tu-dortmund.de}
\author[a]{Daniel Pidt}
\emailAdd{daniel.pidt@tu-dortmund.de}
\affiliation[a]{Fakultät für Physik, TU Dortmund, Otto-Hahn-Str.4, D-44221 Dortmund, Germany}

\abstract{
The ultraviolet completion of flavour models can strongly improve the predictivity of the respective effective models. We consider $A_4$ models, existing minimal UV completions and construct several  next-to-minimal UV complete models. We compare the results of these possibilities to the experimental data including $\theta_{13}$. Through the predictive power of the UV completions, we are able to either rule out or constrain several minimal and next-to-minimal alternatives.
}

\keywords{Flavour symmetries, Fermion masses and mixing, UV completions}

\begin{document}
\maketitle


\section{Introduction}

The ultraviolet (UV) completion of effective models requires a number of mediator fields
that are integrated out below the cut-off typically given by the mediator mass scale, to obtain the effective model.
These mediators are typically Froggatt-Nielsen (FN) messenger fields \cite{Froggatt:1978nt}, although \cite{Varzielas:2012ss} presents a novel possibility using Higgs mediators.
As shown explicitly in \cite{Varzielas:2010mp}, UV completions of flavor models tend to have a subsection of all possible next-to-leading order (NLO) terms of the respective effective theories. This is particularly so when these completions are minimal, in the sense of having the smallest set of mediator fields and associated renormalisable terms that enable the desired leading order (LO) features of the respective effective model.
In \cite{Varzielas:2010mp}, minimal completions of $A_4$ models \cite{Altarelli:2005yx} (AF) and \cite{Altarelli:2009kr} (AM) were presented. The effective models in question predict tri-bi-maximal (TB) mixing, which was in good agreement with neutrino data at the time. With the measurement of $\theta_{13}$ (see the global fits \cite{Tortola:2012te, *Fogli:2012ua, *GonzalezGarcia:2012sz} and references therein) exact TB mixing is now excluded, but it remains a favourable starting point to explain the observed mixing. Deviations from the exact TB values must then arise from one or more sectors: the charged lepton (e.g. \cite{King:2012vj, Antusch:2012fb}), Dirac (e.g. \cite{deMedeirosVarzielas:2011wx}) or Majorana (in type I seesaw models, as in \cite{Varzielas:2012ss}), or vacuum expectation value (VEV) alignment (e.g. \cite{King:2009qt}). For a recent review, see e.g. \cite{Altarelli:2012ss}.

The $A_{4}$ group structure does not necessary lead to TB mixing (see e.g. \cite{Ma:2004zv}). Large $\theta_{13}$ can be obtained at LO and this can be realised by having flavons transforming as $1'$ or $1''$ under $A_4$ \cite{Brahmachari:2008fn} (see also \cite{Barry:2010zk} and more recently \cite{Shimizu:2011xg}).
Other recent works using $A_4$ and obtaining large $\theta_{13}$ include \cite{Ahn:2012tv} (with an extended $SU(2)$ doublet sector) and \cite{Chen:2012st}. In \cite{Chen:2012st} deviations from TB mixing are explored in an analysis where $\theta_{12}$ is kept fixed to the TB value, which we note is not the case in the models we consider here.

In terms of the UV complete models presented in \cite{Varzielas:2010mp}, it is trivial to conclude that the minimal completion of the AF model is decidedly excluded by the observation of non-zero $\theta_{13}$, it leads to exact TB mixing. This is in contrast with the ambiguity at the NLO of the predictions of effective models, and also the AM minimal completion which has a specific source of TB deviations.
In section \ref{sec:AF} we consider some possible next-to-minimal completions of the AF model which only add a few extra messengers, and compare that approach to modifications changing instead the flavon content. The latter approach lead us to a model that could also be obtained by providing a minimal UV completion of the type of effective model discussed recently in \cite{Shimizu:2011xg}.
In section \ref{sec:AM} we address first the minimal completion of the AM model presented in \cite{Varzielas:2010mp}. It does not predict exact TB mixing, but the minimal completion is still strongly predictive and we find it is quite constrained by the present experimental values of the mixing angles. We then explore different possibilities of next-to-minimal completions in some detail. Finally, as with the AF case, we compare with an approach where instead the flavon content is modified.

We follow the same conventions in terms of $A_{4}$ transformations as in \cite{Varzielas:2010mp}. In terms of notation, we use curly brackets to indicate $A_{4}$ products: for two $A_{4}$ triplets $A$ and $B$, $\left\{ A B \right\}=(A_1 B_1 + A_2 B_3 + A_3 B_2) \sim \mathbf{1}$, $\left\{ A B \right\}' =(A_1 B_2 + A_2 B_1 + A_3 B_3) \sim \mathbf{1}'$ and $\left\{ A B \right\}'' = (A_1 B_3 + A_2 B_2 + A_3 B_1)\sim \mathbf{1}''$ and similarly for three triplet contractions. We have also renamed the triplet flavon fields to $\phi_l$ and $\phi_\nu$ (formerly $\phi_T$ and $\phi_S$ respectively).

\section{$A_{4}\times Z_{3}\times U\left( 1 \right)_{\textrm{FN}}$ models \label{sec:AF}}

The basis of the models discussed in this section is the supersymmetric (SUSY) implementation of the AF model \cite{Altarelli:2005yx}. Its flavor symmetry is given by the product $A_{4}\times Z_{3}\times U\left( 1 \right)_{\textrm{FN}}$. The original SUSY AF model produces TB leptonic mixing at the leading order, through the spontaneous breaking of $A_{4}$. The $Z_{3}$ separates the neutrino sector and the charged lepton sector and prevents dangerous couplings, while the FN mechanism \cite{Froggatt:1978nt} is implemented separately through a traditional $U\left( 1 \right)_{\textrm{FN}}$ that generates the hierarchy in the charged lepton masses naturally.

The minimal completion of the AF model presented in \cite{Varzielas:2010mp} is elegant and has the rather unique feature of predicting exact TB mixing. The field content is presented in table \ref{tab:assignment} and table \ref{tab:UVassignment} (as discussed later in this section the field $\tilde{\xi}$ in table \ref{tab:assignment} is only present in the original AF model and its minimal completion from \cite{Varzielas:2010mp}, and it is replaced by $\xi'$ in the new minimal model that we present here). Clearly, the recent experimental evidence excludes the original minimal completion. What about next-to-minimal completions? Given it is a type I seesaw model, we can consider changes to the VEV alignment, the charged lepton mass terms, the neutrino Majorana or Dirac terms.

\begin{table}
  \centering
\footnotesize{
  \begin{tabular}{|c||ccccc|cc|cccc|cc|ccc|}
    \hline
    & $\nu^{c}$ & $l$ & $e^{c}$ & $\mu^{c}$ & $\tau^{c}$  & $h_{d}$ & $h_{u}$ & $\theta$  & $\phi_{l}$ & $\phi_{\nu}$ & $\xi$ & $\tilde{\xi}$  & $\xi'$ & $\phi_{l}^{0}$ & $\phi_{\nu}^{0}$ & $\xi^{0}$  \\
    \hline\hline
    $A_{4}$ & $\mathbf{3}$  & $\mathbf{3}$  & $\mathbf{1}$  & $\mathbf{1''}$  & $\mathbf{1'}$  & $\mathbf{1}$  & $\mathbf{1}$  & $\mathbf{1}$  & $\mathbf{3}$  & $\mathbf{3}$  & $\mathbf{1}$  & $\mathbf{1}$  & $\mathbf{1'}$  & $\mathbf{3}$  & $\mathbf{3}$  & $\mathbf{1}$ \\
    $Z_{3}$ & $\omega^{2}$  & $\omega$  & $\omega^{2}$  & $\omega^{2}$  & $\omega^{2}$  & $1$ & $1$ & $1$ & $1$ & $\omega^{2}$  & $\omega^{2}$  & $\omega^{2}$  & $\omega^{2}$ & $1$   & $\omega^{2}$ & $\omega^{2}$  \\
    $U(1)_{\textrm{FN}}$ & $0$ & $0$ & $2$ & $1$ & $0$ & $0$ & $0$ & $-1$ & $0$ & $0$ & $0$ & $0$ & $0$ & $0$ & $0$ &  $0$ \\  
    $U(1)_{\textrm{R}}$ & $1$ & $1$ & $1$ & $1$ & $1$ & $0$ & $0$ & $0$ & $0$ & $0$ & $0$ & $0$ & $0$ & $2$ & $2$ &  $2$ \\  
    $U(1)_{\textrm{Y}}$ & $0$ & $-1/2$ & $+1$ & $+1$ & $+1$ & $-1/2$ & $+1/2$ & $0$ & $0$ & $0$ & $0$ & $0$ & $0$ & $0$ & $0$  & $0$ \\  
    \hline
  \end{tabular}
   \caption{Field assignment of the AF model and its UV completions. The original AF model \cite{Altarelli:2005yx} and minimal completion \cite{Varzielas:2010mp} do not contain $\xi'$, while the new completion we propose does not have $\tilde{\xi}$. \label{tab:assignment}}
}
  
\end{table}

\begin{table}
  \centering
  \begin{tabular}{|c||cccc|cccc|}
    \hline
    & $\chi_{\tau}$ & $\chi_{1}$ & $\chi_{2}$ & $\chi_{3}$  & $\chi_{\tau}^{c}$ & $\chi_{1}^{c}$ & $\chi_{2}^{c}$ & $\chi_{3}^{c}$  \\
    \hline\hline
    $A_{4}$ & $\mathbf{3}$  & $\mathbf{1'}$  & $\mathbf{1}$  & $\mathbf{1}$  & $\mathbf{3}$  & $\mathbf{1''}$ & $\mathbf{1}$  & $\mathbf{1}$  \\
    $Z_{3}$ & $\omega$  & $\omega$  & $\omega$  & $\omega$  & $\omega^{2}$  & $\omega^{2}$ & $\omega^{2}$ & $\omega^{2}$ \\
    $U(1)_{\textrm{FN}}$ & $0$ & $0$ & $0$ & $-1$ & $0$ & $0$ & $0$ & $+1$  \\
    $U(1)_{\textrm{R}}$ & $1$ & $1$ & $1$ & $1$ & $1$ & $1$ & $1$ & $1$  \\
    $U(1)_{\textrm{Y}}$ & $-1$ & $-1$ & $-1$ & $-1$ & $+1$ & $+1$ & $+1$ & $+1$  \\
    \hline
  \end{tabular}
  \caption{The FN messengers remain the same as presented in \cite{Varzielas:2010mp}. \label{tab:UVassignment}}
\end{table}

The model is based on the MSSM, although with several additional superfields. We class them as flavons (gauge singlets with $U(1)_{\textrm{R}}$ R-charge $0$) or alignment fields that have a superscript $0$ (gauge singlets with R-charge $2$). The mediators for the explicit completion are FN messengers and are denoted generically as $\chi$ fields. The messengers carry R-charge $1$, like the leptons.
The superpotential for the original AF model (with $\tilde{\xi}$) is
\begin{align}
  w &=  w_{l} + w_{\nu} + w_{d},
  \label{eqn:AFSPot}
\end{align}
divided into the driving superpotential giving rise to the alignment of flavon VEVs
\begin{align}
  w_{d} &=  M\left\{ \phi^{0}_{l}\phi_{l} \right\}  + g\left\{ \phi_{l}^{0}\phi_{l}\phi_{l} \right\} + g_{1}\left\{ \phi_{\nu}^{0}\phi_{\nu}\phi_{\nu} \right\} + g_{2}\tilde{\xi}\left\{ \phi_{\nu}^{0}\phi_{\nu} \right\} + g_{3}\xi^{0}\left\{ \phi_{\nu}\phi_{\nu} \right\} \nonumber\\
  &\quad  + g_{4}\xi^{0}\xi\xi  + g_{5}\xi^{0}\xi\tilde{\xi}  + g_{6}\xi^{0}\tilde{\xi}\tilde{\xi} ,
\label{eqn:AFVEVSPot}
\end{align}
the neutrino superpotential
\begin{align}
  w_{\nu} &=  y\left\{l v^{c} \right\}h_{u} + \left( x_{A}\xi + \tilde{x}_{A}\tilde{\xi} \right)\left\{ \nu^{c}\nu^{c} \right\} + x_{B}\left\{ \phi_{\nu}\nu^{c}\nu^{c} \right\} ,
  \label{eqn:AFNuSPot}
\end{align}
and the charged lepton superpotential
\begin{align}
  w_{l} &=  \frac{y_{e}}{\Lambda^{3}}\theta^{2}e^{c}\left\{ \phi_{l}l \right\}h_{d}  + \frac{y_{\mu}}{\Lambda^{2}}\theta\mu^{2}\left\{ \phi_{l}l \right\}'h_{d}  + \frac{y_{\tau}}{\Lambda}\tau^{c}\left\{ \phi_{l}l \right\}''h_{d} .
  \label{eqn:AFCLepSPot}
\end{align}
The renormalisable charged lepton superpotential of the minimal completion giving rise to this effective potential is
\begin{align}
  w_{l}^{\mathrm{UV}} &=  M_{\chi_{A}}\left\{ \chi_{A}\chi_{A}^{c} \right\} + h_{d}\left\{ l\chi_{\tau}^{c} \right\} + \tau{^c}\left\{ \phi_{l}\chi_{\tau} \right\}'' + \theta\mu^{c}\chi_{1} + \theta e^{c}\chi_{3} \nonumber \\
 &\quad+ \left\{ \phi_{l}\chi_{\tau} \right\}'\chi_{1}^{c}  + \left\{ \phi_{l}\chi_{\tau} \right\}\chi_{2}^{c}  + \theta\chi_{2}\chi_{3}^{c}.
  \label{eqn:AFCLepUVSPot}
\end{align}
The subscript $A$ in the mass term labels the different messenger pairs (see table \ref{tab:UVassignment}), which are expected to have similar masses, denoted generically as $M_\chi$.
We recall that the curly brackets represent the $A_{4}$ contractions $\left\{ a \right\}\sim \mathbf{1}$, $\left\{ a \right\}'\sim \mathbf{1}'$ and $\left\{ a \right\}''\sim \mathbf{1}''$.
The alignment conditions derived from Eq.(\ref{eqn:AFVEVSPot}) are
\begin{align}
  \frac{\partial w}{\partial\phi_{\nu1}^{0}} &=  g_{2}\tilde{\xi}\phi_{\nu1} + \frac{2}{3}g_{1}\left( \phi_{\nu1}^{2} - \phi_{\nu2}\phi_{\nu3} \right)  = 0, \label{eqn:MinimizationAFStart}\\
  \frac{\partial w}{\partial\phi_{\nu2}^{0}} &=  g_{2}\tilde{\xi}\phi_{\nu3} + \frac{2}{3}g_{1}\left( \phi_{\nu2}^{2} - \phi_{\nu1}\phi_{\nu3} \right)  = 0, \\
  \frac{\partial w}{\partial\phi_{\nu3}^{0}} &=  g_{2}\tilde{\xi}\phi_{\nu2} + \frac{2}{3}g_{1}\left( \phi_{\nu3}^{2} - \phi_{\nu1}\phi_{\nu2} \right)  = 0, \\
  \frac{\partial w}{\partial\xi^{0}} &=  g_{3}\left( \phi_{\nu1}^{2}  + 2\phi_{\nu2} \phi_{\nu3} \right)  + g_{4}\xi^{2} + g_{5}\xi \tilde{\xi}  + g_{6} \tilde{\xi}^{2}  = 0,
  \label{eqn:MinimizationAFEnd}
\end{align}
and
\begin{align}
  \frac{\partial w}{\partial\phi_{l1}^{0}} &=  M\phi_{l1} + \frac{2}{3}g_{1}\left( \phi_{l1}^{2} - \phi_{l2}\phi_{l3} \right)  = 0, \label{eqn:MinimizationAFChiLStart}\\
  \frac{\partial w}{\partial\phi_{l2}^{0}} &=  M\phi_{l3} + \frac{2}{3}g_{1}\left( \phi_{l2}^{2} - \phi_{l1}\phi_{l3} \right)  = 0, \\
  \frac{\partial w}{\partial\phi_{l3}^{0}} &=  M\phi_{l2} + \frac{2}{3}g_{1}\left( \phi_{l3}^{2} - \phi_{l1}\phi_{l2} \right)  = 0,
  \label{eqn:MinimizationAFChiLEnd}
\end{align}
which lead to the VEV structure
\begin{align}
 \left<\phi_{\nu}\right> &\propto (1,1,1), \quad  \left<\phi_{l}\right> \propto (1,0,0),  \quad \left<\xi\right> \neq 0, \quad \langle\tilde{\xi}\rangle = 0. 
\end{align}

\begin{figure}
\centering
\includegraphics[width=6 cm]{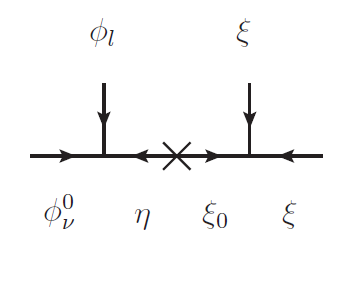}
\caption{Topology for $(\phi_0^\nu \phi_l \xi \xi)$. \label{fig:top}}
\end{figure}

Changing the terms that contribute to the alignment of the VEVs is non-trivial as it can easily lead to trivially vanishing VEVs or other alignments that are not phenomenologically viable. As an example, a simple possibility is enabling $\phi_l$ to appear in the $\phi^0_\nu$ terms. At the non-renormalisable level this occurs with $\left\{ \phi^0_\nu \phi_l \right\} \xi \xi $. The existence of this effective term in a renormalisable theory requires only a new field $\eta$ transforming as $\omega$ under the $Z_3$, c.f. table \ref{tab:assignment}, as $\eta$ enables a mass term $\xi^0 \eta$ and the vertex $\left\{ \phi^0_\nu \phi_l \right\} \eta$. This vertex by itself can already introduce $\phi_l$ mixing into the $\phi^0_\nu$ alignment, but only if $\eta$ acquires a non-vanishing VEV. Yet, even with $\langle \eta \rangle = 0$ the new allowed terms combine with $\xi^0 \xi \xi$ and lead to the effective $\left\{ \phi^0_\nu \phi_l \right\} \xi \xi$ term through the diagram in figure \ref{fig:top}. The existing alignment equations (c.f. Eqs.(\ref{eqn:MinimizationAFStart})-(\ref{eqn:MinimizationAFEnd})) are modified to:
\begin{align}
  \frac{\partial w}{\partial\phi_{\nu1}^{0}} &=  g_{2}\tilde{\xi}\phi_{\nu1} + \frac{2}{3}g_{1}\left( \phi_{\nu1}^{2} - \phi_{\nu2}\phi_{\nu3} \right) + g_{8}\phi_{l1}\eta = 0, \label{eqn:MinimizationAFModifiedStart}\\
  \frac{\partial w}{\partial\phi_{\nu2}^{0}} &=  g_{2}\tilde{\xi}\phi_{\nu3} + \frac{2}{3}g_{1}\left( \phi_{\nu2}^{2} - \phi_{\nu1}\phi_{\nu3} \right)  + g_{8}\phi_{l3}\eta = 0, \\
  \frac{\partial w}{\partial\phi_{\nu3}^{0}} &=  g_{2}\tilde{\xi}\phi_{\nu2} + \frac{2}{3}g_{1}\left( \phi_{\nu3}^{2} - \phi_{\nu1}\phi_{\nu2} \right)  + g_{8}\phi_{l2}\eta = 0, \\
  \frac{\partial w}{\partial\xi^{0}} &=  g_{3}\left( \phi_{\nu1}^{2}  + 2\phi_{\nu2}\phi_{\nu3} \right)  + g_{4}\xi^{2} + g_{5}\xi \tilde{\xi}  + g_{6} \tilde{\xi}^{2} + g_{7}\eta = 0.
  \label{eqn:MinimizationAFModifiedEnd}
\end{align}
It is easy to see $\langle \phi_l \rangle \propto (1,0,0)$ and $\langle \phi_\nu \rangle \propto (1,1,1)$ do not satisfy these modified alignment conditions. Indeed, preserving the alignment of $\left<\phi_{\nu}\right>$, one would require $\left<\phi_{l}\right>\propto\left( 1,1,1 \right)$ which is not compatible with Eqs.(\ref{eqn:MinimizationAFChiLStart})-(\ref{eqn:MinimizationAFChiLEnd}). This problem can not be remedied by allowing small perturbations of the structure of $\left< \varphi_{\nu} \right>$.
Modifications to the alignment sector will be considered in more detail in section \ref{sec:AM}, within the $A_{4} \times Z_4$ framework.

The charged lepton effective terms are already at the non-renormalisable level and are not a promising source for deviations that enable a large $\theta_{13}$. The effect of the renormalisable vertex $\left\{ \chi_\tau^c \chi_\tau \phi_l \right\}$ is already discussed in \cite{Varzielas:2010mp} as not affecting the leptonic mixing, and the non-renormalisable term $\chi_\tau^c \chi_\tau \phi_l \phi_l$ appears through a diagram that merely involves that vertex twice.

\begin{figure}
\centering
\includegraphics[width=6 cm]{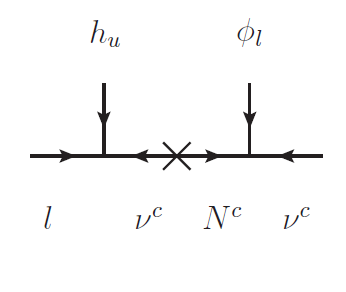}
\caption{Diagram contributing to $m_D$, enabled by $N^{c}$. \label{fig:AFmD}}
\end{figure}

Extending the Majorana sector requires adding R-charge $1$ fields with vanishing hypercharge. One possible non-renormalisable term is $\nu^c \phi_\nu \phi_l \nu^c$, with field $N^c$ transforming as $\omega$ under $Z_3$, enabling the mass term $\nu^c N^c$ and vertex $\left\{ \nu^c \phi_l N^c \right\}$. This amounts to an extended seesaw realisation, and one can see that the effective Dirac mass term $l h_u \phi_l \nu^c$ (allowed by $A_{4}\times Z_{3}\times U\left( 1 \right)_{\textrm{FN}}$, but not present in the minimal completion) arises from the diagram in figure \ref{fig:AFmD}.
The contribution to $m_D$ is of the form
\begin{align}
  \delta m  &=  
  \begin{pmatrix}
    2a  & 0 & 0 \\
    0 & b-a & 0 \\
    0 & 0 & -b-a
  \end{pmatrix} .
  \label{eqn:AFNuDiracCorrection}
\end{align}
Without fine-tuning the parameters $a$,$b$, $\delta m$ leads to significant changes of $\theta_{12}$ while not affecting the other angles as significantly (note the $a$ contribution preserves $\mu$-$\tau$ symmetry). It is therefore not viable to obtain a large value of $\theta_{13}$ by adding $N^{c}$.
It is also possible to obtain the effective $l h_u \phi_l \nu^c$ Dirac term by introducing an $SU(2)$ doublet messenger (the same messenger would also lead to new contributions to the charged lepton masses with an additional $\phi_l$ insertion, e.g. $l \phi_l h_d \phi_l \tau^c$, but those would merely redefine existing terms).

At this stage we conclude that next-to-minimal completions of the AF model are not very successful in generating large $\theta_{13}$. One can further consider completions where a multitude of additional effective terms are enabled by many extra mediators, but this sacrifices predictivity, which is an important motivation for having an explicit completion.
We find it more attractive to replace $\tilde{\xi}$ (a duplicate flavon with the same assignment as $\xi$ that was required to obtain the desired VEV alignment) with the non-trivial $A_{4}$ singlet $\xi'$. The resulting minimal model has the same number of fields a the minimal AF completion and is very similar to the model one would obtain from starting with the different effective model presented in \cite{Shimizu:2011xg} and providing an explicit minimal completion of it.
The full symmetry and field content of the $\xi'$ completion was already summarised in table \ref{tab:assignment}, with the messengers in table \ref{tab:UVassignment}.

The superpotential is modified with
\begin{align}
  w_{d} &=  M\left\{ \phi^{0}_{l}\phi_{l} \right\}  + g\left\{ \phi_{l}^{0}\phi_{l}\phi_{l} \right\} + g_{1}\left\{ \phi_{\nu}^{0}\phi_{\nu}\phi_{\nu} \right\} + g_{2}\xi\left\{ \phi_{\nu}^{0}\phi_{\nu} \right\}  \nonumber\\
  &\quad  + g_{3}\xi'\left\{ \phi_{\nu}^{0}\phi_{\nu} \right\}''  + g_{4}\xi^{0}\left\{ \phi_{\nu}\phi_{\nu} \right\}  + g_{5}\xi^{0}\xi\xi
\label{eqn:AFVEVSPotMod}
\end{align}
and
\begin{align}
  w_{\nu} &=  y\left\{ v^{c}l \right\}h_{u} + \left( x_{A}\xi + x'_{A}\xi' \right)\left\{ \nu^{c}\nu^{c} \right\} + x_{B}\left\{ \phi_{\nu}\nu^{c}\nu^{c} \right\} ,
  \label{eqn:AFNuSPotMod}
\end{align}
replacing Eqs.(\ref{eqn:AFVEVSPot}) and (\ref{eqn:AFNuSPot}) while Eqs.(\ref{eqn:AFCLepSPot}) and (\ref{eqn:AFCLepUVSPot}) remain unchanged.
After symmetry breaking the model leads to a diagonal charged lepton mass matrix as in the original models, as $\xi'$ does not appear in the respective terms (only in the alignment and neutrino terms).
In contrast with \cite{Varzielas:2010mp} and \cite{Altarelli:2005yx}, the extra flavon $\tilde{\xi}$ with the quantum numbers of $\xi$ is absent. As $\tilde{\xi}$ was necessary to obtain a nontrivial VEV structure from the minimisation of the potential in the original models, we must reconsider the minimisation of the potential. Eqs.(\ref{eqn:MinimizationAFChiLStart})-(\ref{eqn:MinimizationAFChiLEnd}) still apply to this model, and we have also:
\begin{align}
  \frac{\partial w}{\partial\phi_{\nu1}^{0}} &=  g_{2}\xi\phi_{\nu1} + g_{3}\xi'\phi_{\nu3} + \frac{2}{3}g_{1}\left( \phi_{\nu1}^{2} - \phi_{\nu2}\phi_{\nu3} \right)  = 0, \label{eqn:MinimizationAFModStart}\\
  \frac{\partial w}{\partial\phi_{\nu2}^{0}} &=  g_{2}\xi\phi_{\nu3} + g_{3}\xi'\phi_{\nu2} + \frac{2}{3}g_{1}\left( \phi_{\nu2}^{2} - \phi_{\nu1}\phi_{\nu3} \right)  = 0, \\
  \frac{\partial w}{\partial\phi_{\nu3}^{0}} &=  g_{2}\xi\phi_{\nu2} + g_{3}\xi'\phi_{\nu1} + \frac{2}{3}g_{1}\left( \phi_{\nu3}^{2} - \phi_{\nu1}\phi_{\nu2} \right)  = 0, \\
  \frac{\partial w}{\partial\xi^{0}} &=  g_{4}\left( \phi_{\nu1}^{2}  + 2\phi_{\nu2} \phi_{\nu3} \right) + g_{5}\xi^{2}.
  \label{eqn:MinimizationAFModEnd}
\end{align}
Note the effect of the $\xi'$ flavon (c.f. Eqs.(\ref{eqn:MinimizationAFStart})-(\ref{eqn:MinimizationAFEnd})).  The VEV structure obtained is very similar to that of the original models:
\begin{align}
  \frac{\left< \phi_{l} \right>}{M_{\chi}}  &=  \left( u,0,0 \right), \quad \frac{\left< \phi_{\nu} \right>}{M_{\chi}} = c_{b}\left( u,u,u \right), \quad  \frac{\left< \xi \right>}{M_{\chi}}  = c_{a}u,  \quad \frac{\left< \xi' \right>}{M_{\chi}} = c_{a}'u ,
  \label{eqn:VEVs}
\end{align}
\begin{align}
  u &=  -\frac{3}{2}\frac{M}{g},  \quad c_{b}^{2} = -\frac{g_{5}g_{3}^{2}}{3g_{4}g_{2}^{2}}c_{a}'^{2}, \quad c_{a} = -\frac{g_{3}}{g_{2}}c_{a}'.
  \label{eqn:udef}
\end{align}
$M_{\chi}$ generically denotes the messenger masses and the parameter $c_{a}'$ remains undetermined. Small perturbations of this VEV structure are incompatible with the alignment equations. 

As the general VEV structure is not modified, what are the changes enabled by $\xi'$? The Dirac mass structure remains the same as in \cite{Altarelli:2005yx}:
\begin{align}
  M_{D} &=  yv_{u}
  \begin{pmatrix}
    1 & 0 & 0 \\
    0 & 0 & 1 \\
    0 & 1 & 0
  \end{pmatrix},
  \label{eqn:MNuD}
\end{align}
where $v_{u}$ denotes the VEV $\left< h_{u} \right>$. On the other hand, due to the new terms in Eq.(\ref{eqn:AFNuSPotMod}) the Majorana mass reads
\begin{align}
  M_{M} &=  x_{A}c_{a}uM_{\chi}
  \begin{pmatrix}
    1 & 0 & 0 \\
    0 & 0 & 1 \\
    0 & 1 & 0
  \end{pmatrix}
  + x_{A}'c_{a}'uM_{\chi}
  \begin{pmatrix}
    0 & 0 & 1 \\
    0 & 1 & 0 \\
    1 & 0 & 0
  \end{pmatrix}
  + \frac{1}{3}x_{B}c_{b}uM_{\chi}
  \begin{pmatrix}
    2 & -1  & -1  \\
    -1  & 2 & -1  \\
    -1  & -1  & 2
  \end{pmatrix}.
  \label{eqn:MNuM}
\end{align}
Here, the second term is related to the new flavon $\xi'$ and it is this structure in the Majorana mass that will enable large $\theta_{13}$. Indeed one can check that it leads to the effective neutrino mass matrix presented in \cite{Shimizu:2011xg}
\begin{align}
  M_{\nu} &=  a 
  \begin{pmatrix}
    1 & 0 & 0 \\
    0 & 1 & 0 \\
    0 & 0 & 1
  \end{pmatrix}
  + b
  \begin{pmatrix}
    1 & 1 & 1 \\
    1 & 1 & 1 \\
    1 & 1 & 1
  \end{pmatrix}
  +c
  \begin{pmatrix}
    1 & 0 & 0 \\
    0 & 0 & 1 \\
    0 & 1 & 0
  \end{pmatrix}
  +d
  \begin{pmatrix}
    0 & 0 & 1 \\
    0 & 1 & 0 \\
    1 & 0 & 0
  \end{pmatrix}
  \label{eqn:MNuStructure}
\end{align}
with the effective parameters related to those that appear in the Dirac and Majorana mass matrices
\begin{align}
  a &=  \frac{y^{2}v_{u}^{2}}{M_{\chi} u}\frac{x_{B}c_{b}}{-x_{A}^{2}c_{a}^{2} + x_{A}c_{a}x_{A}'c_{a}' - x_{A}'^{2}c_{a}'^{2}  + x_{B}^{2}c_{b}^{2}}, \\
  b &=  \frac{y^{2}v_{u}^{2}}{3M_{\chi} u}\frac{3x_{A}'^{2}c_{a}'^{2}  + \left( x_{A}c_{a} + x_{A}'c_{a}' \right)x_{B}c_{b} + x_{B}^{2}c_{B}^{2}}{x_{A}^{3}c_{a}^{3}  + x_{A}'^{3}c_{a}'^{3}-x_{B}^{2}c_{b}^{2}\left( x_{A}c_{a}  + x_{A}'c_{a}' \right)}, \\
  c &=  \frac{y^{2}v_{u}^{2}}{M_{\chi} u}\frac{x_{A}c_{a}-x_{A}'c_{a}'}{x_{A}^{2}c_{a}^{2}-x_{A}c_{a}x_{A}'c_{a}' + x_{A}'^{2}c_{a}'^{2}-x_{B}^{2}c_{b}^{2}}, \\
  d &=  -\frac{y^{2}v_{u}^{2}}{M_{\chi} u}\frac{x_{A}'c_{a}'}{x_{A}^{2}c_{a}^{2}-x_{A}c_{a}x_{A}'c_{a}' + x_{A}'^{2}c_{a}'^{2}-x_{B}^{2}c_{b}^{2}}.
  \label{eqn:translate}
\end{align}
For this structure the matrix diagonalising $M_{\nu}$ can be expressed by an additional rotation applied to the TB mixing matrix $V_{\textrm{TBM}}$:
\begin{align}
  U_{\textrm{PMNS}} &=  V_{\textrm{TBM}}
  \begin{pmatrix}
    \cos\theta  & 0 & \sin\theta  \\
    0 & 1 & 0 \\
    -\sin\theta & 0 & \cos\theta  \\
  \end{pmatrix} ,
  \label{eqn:UPMNS}
\end{align}
with
\begin{align}
  V_{\textrm{TBM}}  &=
  \begin{pmatrix}
    \frac{2}{\sqrt{6}}  & \frac{1}{\sqrt{3}}  & 0 \\
    \frac{-1}{\sqrt{6}} & \frac{1}{\sqrt{3}}  & \frac{-1}{\sqrt{2}} \\
    \frac{-1}{\sqrt{6}} & \frac{1}{\sqrt{3}}  & \frac{1}{\sqrt{2}} 
  \end{pmatrix}
  \label{eqn:VTBM}
\end{align}
and
\begin{align}
  \tan2\theta  &= \frac{\sqrt{3}d}{-2c  + d}.
  \label{eqn:DefTheta}
\end{align}
It is clear that $d=0$ corresponds to TB mixing. In that limit it is well known that $M_{\nu}$ is invariant under $Z_2 \times Z_2$ symmetries with well defined matrices in the flavour basis, one of these $Z_2$ corresponding to the $\mu-\tau$ symmetry. Adding the last term in eq.(\ref{eqn:MNuStructure}) $M_{\nu}$ is no longer invariant under the $\mu-\tau$ symmetry, although it remains invariant under the other $Z_2$ associated with TB mixing - this type of situation with one residual $Z_2$ symmetry was discussed recently in \cite{Hernandez:2012ra}. In the case discussed here,
the model generates trimaximal mixing \cite{Haba:2006dz,*He:2006qd,*Grimus:2008tt,*Ishimori:2010fs}
\begin{align}
  \left| U_{e2} \right| &=  \frac{1}{\sqrt{3}},\quad \left| U_{e3} \right|  = \frac{2}{\sqrt{6}}\left| \sin{\theta} \right|,\quad \left| U_{\mu3} \right| = \left| -\frac{1}{\sqrt{6}}\sin{\theta}  - \frac{1}{\sqrt{2}}\cos{\theta} \right| .
  \label{eqn:PMNSElems}
\end{align}

\begin{figure}[t]
  \begin{center}
    \includegraphics[width=1.0\textwidth]{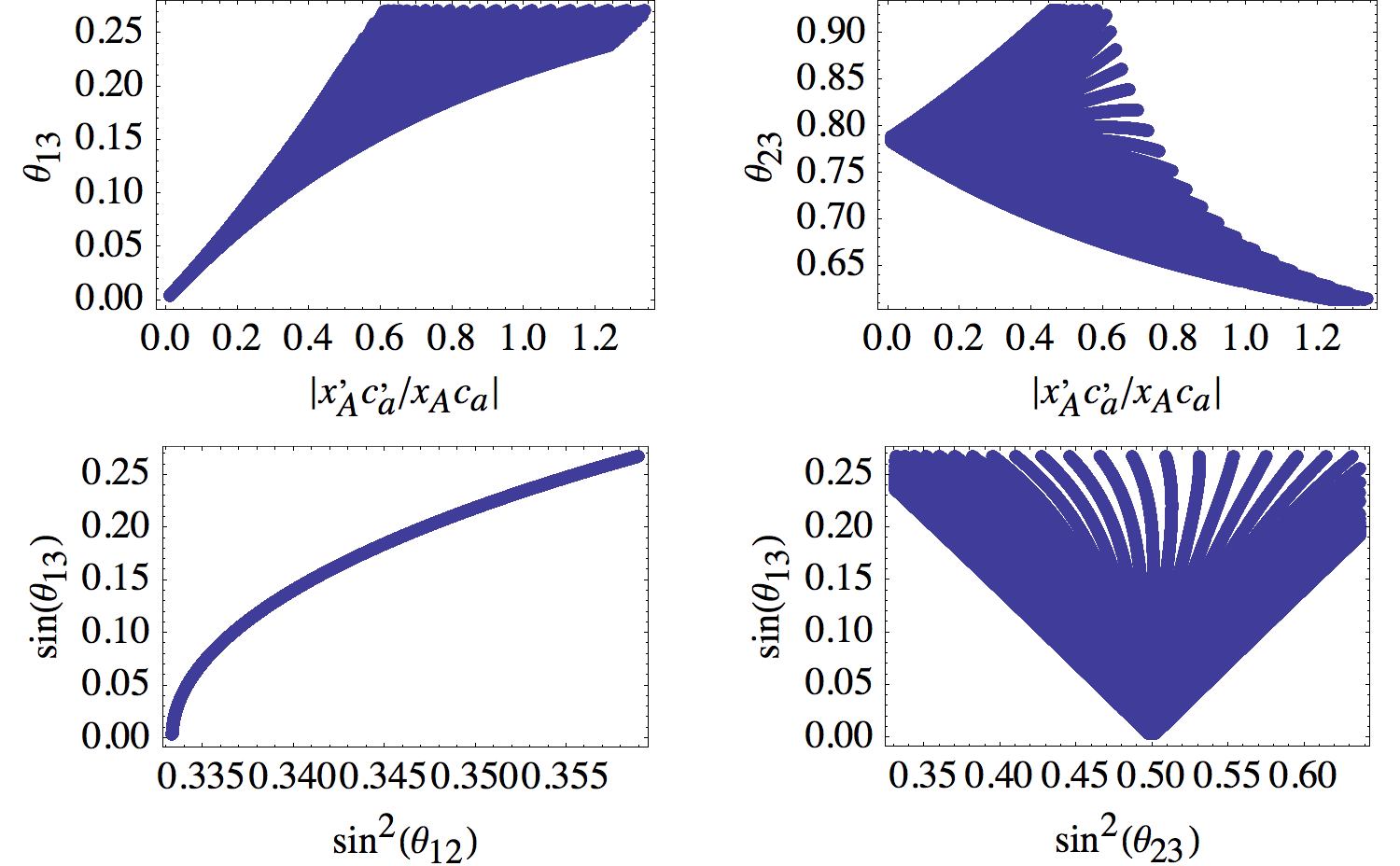}
  \end{center}
  \caption{Deviations of $\theta_{13}$ and $\theta_{23}$ from TB values for complex parameters (top row). Correlation of $\theta_{13}$ to $\theta_{12}$ and $\theta_{23}$ (bottom row). }
  \label{fig:AFPlots}
\end{figure}
We present in figure~\ref{fig:AFPlots} the angles as a function of the parameters appearing in Eq.(\ref{eqn:MNuM}). Our findings agree with the LO results presented in \cite{Shimizu:2011xg}, although we emphasise that the model presented here is renormalisable with an explicit UV completion, including type I seesaw and FN messengers. As such, in this renormalisable model the predictions illustrated in figure~\ref{fig:AFPlots} are not just the LO values.
Furthermore, the model we present here is remarkably simple, having the same symmetry content and number of fields as the minimal AF completion presented in \cite{Varzielas:2010mp}. We have shown that the new complete model leads to viable VEV alignment and to desirable structures for the charged leptons and neutrinos.

Finally, within this framework it is interesting to briefly consider what happens to the effective parameters $a,b,c,d$ when an additional flavon transforming as $1''$ is introduced. At the effective level it is easy to see that it is superfluous \cite{Shimizu:2011xg}. The structure associated with the $1''$ flavon is
\begin{align}
  \begin{pmatrix}
    0 & 1 & 0 \\
    1 & 0 & 0 \\
    0 & 0 & 1
  \end{pmatrix}
  \label{eqn:MPP}
\end{align}
and can simply be absorbed in a redefinition of $a,b,c,d$ as the respective structures are not linearly independent. In fact
\begin{align}
  \begin{pmatrix}
    0 & 1 & 0 \\
    1 & 0 & 0 \\
    0 & 0 & 1
  \end{pmatrix}
  &=  
  \begin{pmatrix}
    1 & 1 & 1 \\
    1 & 1 & 1 \\
    1 & 1 & 1
  \end{pmatrix}
  -
  \begin{pmatrix}
    1 & 0 & 0 \\
    0 & 0 & 1 \\
    0 & 1 & 0
  \end{pmatrix}
  -
  \begin{pmatrix}
    0 & 0 & 1 \\
    0 & 1 & 0 \\
    1 & 0 & 0
  \end{pmatrix}.
  \label{}
\end{align}
If one adds a $1''$ field to the model we present here though, it is the Majorana structure in Eq.(\ref{eqn:MNuM}) that is modified by adding a new coefficient $x_A''$ and the structure in Eq.(\ref{eqn:MPP}). This can be reabsorbed without loss of generality by redefinition of $x_A c_a \rightarrow x_A c_a -x_A'' c_a''$, $x_A' c_a' \rightarrow x_A' c_a' -x_A'' c_a''$, $x_B c_b \rightarrow x_B c_b -3 x_A'' c_a''$ and adding a contribution proportional to the identity matrix (which shifts the overall mass scale by $3 x_A'' c_a''$ but does not affect the mixing). But as seen in Eqs.(\ref{eqn:translate}), due to the seesaw the effective parameters $a,b,c,d$ are related through non-linear expressions to the coefficients that appear in the Majorana mass matrix. For this reason it is convenient to consider the ratio $d/c = x_A' c_a'/(x_A c_a - x_A' c_a')$, and as $x_A c_a$ and $x_A' c_a'$ both get redefined by $x_A'' c_a''$, the effect of the redefinition translates into a linear change of $d/c$ through the shift in the redefined $x_A' c_a'$.

\section{$A_{4} \times Z_4$ models \label{sec:AM}}

The framework discussed in this section includes the AM model \cite{Altarelli:2009kr}, a rather simple $A_{4}$ model that separates the charged lepton and neutrino sectors and provides mass hierarchies through the Froggatt-Nielsen mechanism with a single 
$Z_4$ (c.f. $Z_3 \times U\left( 1 \right)_{\textrm{FN}}$ of the framework discussed in section \ref{sec:AF}). Table \ref{tab:Massignment} lists the field and symmetry content of the original AM effective model, and table \ref{tab:MUVassignment} has the messenger content of its minimal completion as proposed in \cite{Varzielas:2010mp}.

\begin{table} [h]
\centering
\begin{tabular}{|c||c|c|c|c|c||c|c||c|c|c|c||c|c|c|c|}
\hline
&&&&&&&&&&&&&&\\[-4mm]
{\tt Field}& $\nu^c$ & $\ell$ & $e^c$ & $\mu^c$ & $\tau^c$ & $h_d$ & $h_u$& 
$\phi_l$ &  $\xi'$ & $\phi_\nu$ & $\xi$ & $\phi_l^0$  & $\phi_\nu^0$ & $\xi_0$\\[2mm]
\hline
&&&&&&&&&&&&&&\\[-4mm]
$A_{4}$ & $3$ & $3$ & $1$ & $1$ & $1$ & $1$ &$1$ &$3$ & $1'$ & $3$ & $1$ &  $3$ &  $3$ & $1$\\[2mm]
$Z_4$ & $-1$ & $i$ & $1$ & $i$ & $-1$ & $1$ & $i$ & $i$ & $i$  & $1$ & $1$ & $-1$ &  $1$ & $1$\\[2mm]
$U(1)_R$ & $1$& $1$ & $1$ & $1$ & $1$ & $0$ & $0$ & $0$& $0$  & $0$ & $0$ & $2$ & $2$ & $2$\\[2mm]
$U(1)_Y$ & $0$& $-^1/_2$ & $+1$ & $+1$ & $+1$ & $-^1/_2$ & $+^1/_2$ & $0$& $0$  & $0$ & $0$ & $2$ & $2$ & $2$\\[2mm]
\hline
\end{tabular}
\caption{The field and content assignment of the original AM model \cite{Altarelli:2009kr}. \label{tab:Massignment}}
\end{table}

\begin{table}[ht]
\begin{center}
\begin{tabular}{|c||cccc||cccc|}
\hline
&&&&&&&&\\[-4mm]
 & $\chi_\tau$ & $\chi_1$ & $\chi_2$ & $\chi_3$ & $\chi^c_\tau$ & $\chi^c_1$ & $\chi^c_2$ & $\chi^c_3$ \\[2mm]
\hline
&&&&&&&&\\[-4mm]
$A_{4}$ & $\bf3$ & $\bf1''$ & $\bf1'$ & $\bf1''$ & $\bf3$ & $\bf1'$ & $\bf1''$ & $\bf1'$ \\[2mm]
$Z_4$ & $i$ & $-1$ & $-1$ & $-i$ & $-i$ & $-1$ & $-1$ & $i$ \\[2mm]
$U(1)_R$ & 1 & 1 & 1 & 1 & 1 & 1 & 1 & 1 \\[2mm]
$U(1)_Y$ & $-1$ & $-1$ & $-1$ & $-1$ & $+1$ & $+1$ & $+1$ & $+1$ \\[2mm]
\hline
\end{tabular}
\end{center}
\caption{The messenger content of a minimal completion \cite{Varzielas:2010mp}. \label{tab:MUVassignment}}
\end{table}

The neutrino superpotential is
\begin{align}
w_\nu = y_\nu \left\{ l \nu^c \right\} h_u + (M+a\xi) \left\{ \nu^c \nu^c \right\} + b \left\{ \phi_\nu\nu^c\nu^c \right\} ,
\label{AM_nu}
\end{align}
containing the terms that lead to the Dirac and Majorana neutrino mass matrices. 
The driving superpotential
\begin{align}
w_d &= M \left\{ \phi_0^\nu \phi_\nu \right\} + g_1 \left\{ \phi_0^\nu \phi_\nu\phi_\nu \right\}  + g_2 \xi \left\{ \phi_0^\nu \phi_\nu \right\}  + g_3 \xi_0 \left\{ \phi_\nu\phi_\nu \right\} +  g_4 \xi_0 \xi^2 + M_\xi \xi_0 \xi \nonumber\\
 & \quad  +  M_0^2 \, \xi_0 +
h_1 \xi' \left\{ \phi_0^l \phi_l \right\}''+
h_2 \left\{ \phi_0^l \phi_l\phi_l \right\} ,
\label{eqn:AM_driving}
\end{align}
gives rise to the alignment equations. For $\phi_\nu$ they act effectively like Eqs.(\ref{eqn:MinimizationAFStart})-(\ref{eqn:MinimizationAFEnd}) and its VEV is aligned in the $(1,1,1)$ direction as in section \ref{sec:AF} (merely replace $g_2 \tilde{\xi}$ with $(M+g_2 \xi)$). For  $\phi_l$ and $\xi'$ we have
\begin{align}
\frac{\partial w}{\partial \phi^l_{01}}=2 h_2({\phi_l}^2_1-{\phi_l}_2\,{\phi_l}_3)+
h_1\,\xi'\,{\phi_l}_3=0 \label{Malign_start} , \\
\frac{\partial w}{\partial \phi^l_{02}}=2 h_2({\phi_l}^2_2-{\phi_l}_1\,{\phi_l}_3)+
h_1\,\xi'\,{\phi_l}_2=0 , \\
\frac{\partial w}{\partial \phi^l_{03}}=2 h_2({\phi_l}^2_3-{\phi_l}_1\,{\phi_l}_2)+
h_1\,\xi'\,{\phi_l}_1=0 .
\label{Malign_end}
\end{align}
The most relevant difference to the previously discussed framework is due to $\xi'$, which aligns $\langle \phi_l \rangle$ in the $(0,1,0)$ direction. The VEVs are then:
\begin{align}
  \frac{\left< \phi_{l} \right>}{M_{\chi}}  &=  \left( 0,u,0 \right), \quad \frac{\left< \phi_{\nu} \right>}{M_{\chi}} = \varepsilon' \left(1,1,1 \right) \,.
  \label{eqn:MVEVs}
\end{align}

\begin{figure}[htbp]
  \begin{center}
    \includegraphics[width=.8\textwidth]{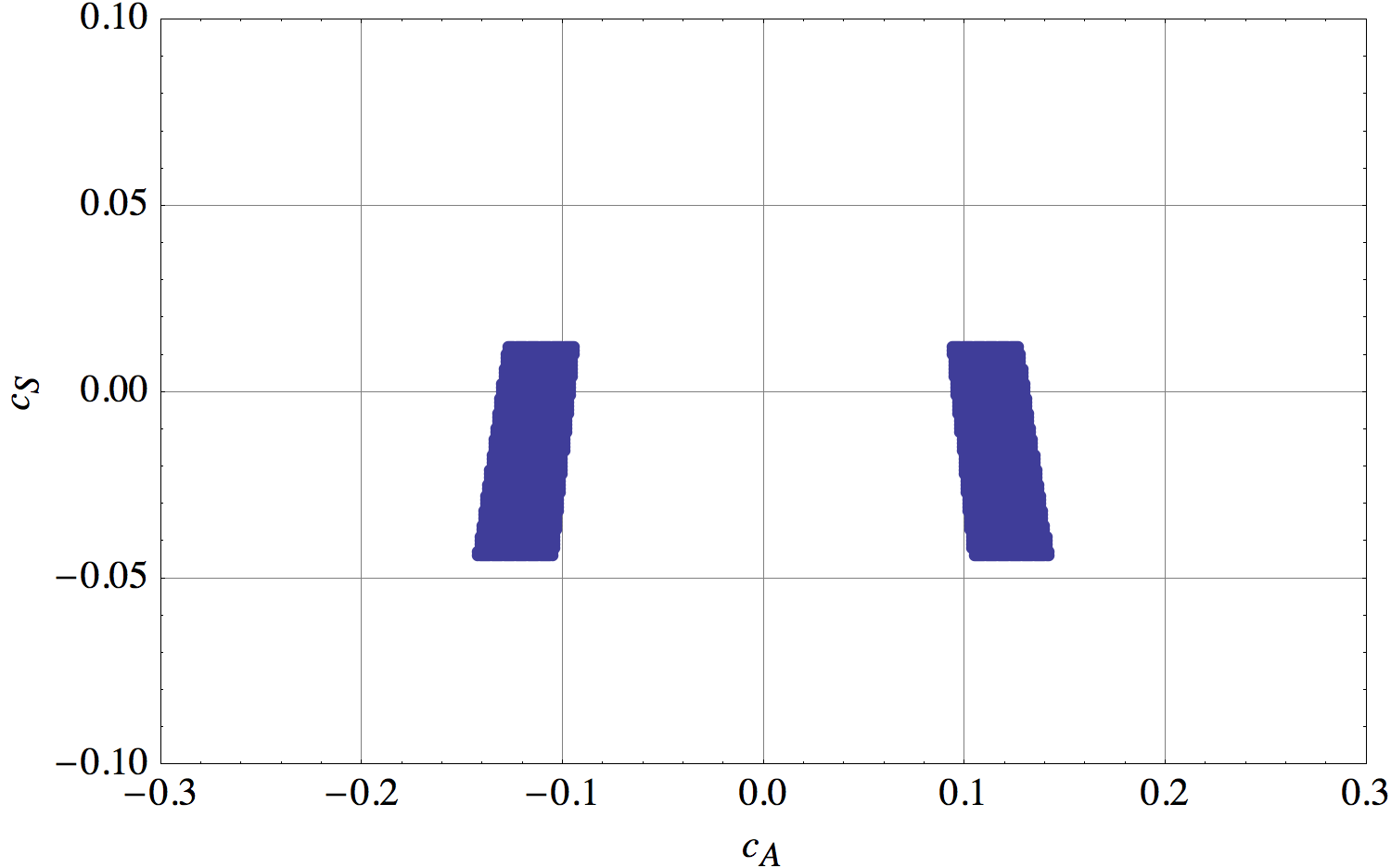}
  \end{center}
  \caption{Range of $c_{A}$ and $c_{S}$ allowed by present mixing angle data for the minimal AM completion \cite{Varzielas:2010mp} ($\varepsilon' = 1$).}
  \label{fig:AM_CAvsCS}
\end{figure}

The minimal completion presented in \cite{Varzielas:2010mp} contains the messengers listed in table \ref{tab:MUVassignment} and already predicts deviations from the TB mixing angles: the minimal messenger content needed to generate the required effective LO terms unavoidably leads also to the term $\left\{ \phi_\nu \chi_\tau \chi^c_\tau \right\}$ which is responsible for making the resulting charged lepton mass matrix non-diagonal:
\begin{equation}
m_{l}=\left(
         \begin{array}{ccc}
           m_e & (-c_s+c_a) \varepsilon' m_\mu & (-c_s-c_a) \varepsilon'  m_\tau \\
           (-c_s+c_a) \varepsilon' m_e & m_\mu & 2 c_s \varepsilon' m_\tau \\
           \varepsilon' (-c_s-c_a) m_e & 2 c_s \varepsilon' m_\mu & m_\tau \\
         \end{array}
       \right)\,,
\label{eqn:AM_mcl}
 \end{equation}
where $\varepsilon'$ is the VEV of $\phi_\nu$ as in Eq.(\ref{eqn:MVEVs}) and $c_a$, $c_s$ are the superpotential parameters governing the $\left\{ \phi_\nu \chi_\tau \chi^c_\tau \right\}$  antisymmetric and symmetric $A_{4}$ invariants respectively.
To a good approximation, the off-diagonal entries are written in terms of the charged lepton masses in order to give a better idea of the relative magnitudes of the entries (see \cite{Varzielas:2010mp}, and note that here we are using a different convention for the mass matrices).
We can find the deviations from TB by diagonalising $m_l m_l^{\dagger}$. In order to generate large contributions to $\theta_{13}$, the $13$ entry of Eq.(\ref{eqn:AM_mcl}) is particularly relevant, but simultaneously the $23$ entry can create undesirable deviations on the other angles that can drive them outside the experimental allowed ranges. Therefore, the allowed region in the $(c_a,c_s)$ parameter space (for a given $\varepsilon'$) is close to the $c_s=0$ axis, with $c_a$ directly constrained by the experimentally allowed values of $\theta_{13}$. This is a clear example of the predictivity of minimal UV complete models: the measurement of the mixing angles is directly probing superpotential parameters of the FN messenger sector. For real $c_a$ and $c_s$ this is illustrated in figure~\ref{fig:AM_CAvsCS} where we take $\varepsilon'=1$ and show the regions that can reproduce the mixing angles within the experimental $3 \sigma$ ranges (for complex parameters the analysis becomes more complicated but the same reasoning remains valid).

\begin{figure}
\centering
	\includegraphics[width=4 cm]{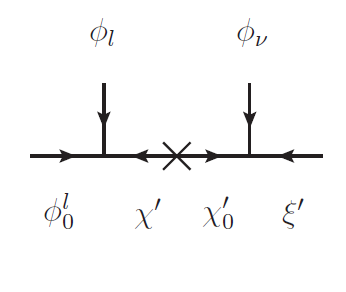}
	\includegraphics[width=4 cm]{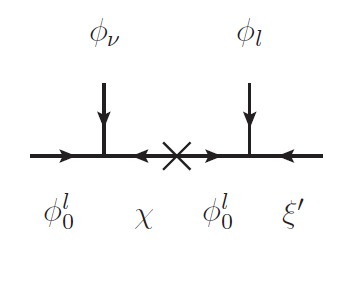}
	\includegraphics[width=4 cm]{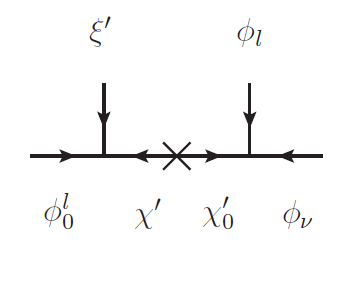}
\caption{Topologies for $\xi' \left\{ \phi_0^l \phi_l \phi_\nu \right\}''$. \label{fig:top_abc}}
\end{figure}

\begin{figure}
\centering
\includegraphics[width=5 cm]{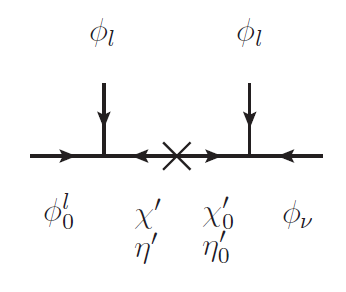}
\includegraphics[width=5 cm]{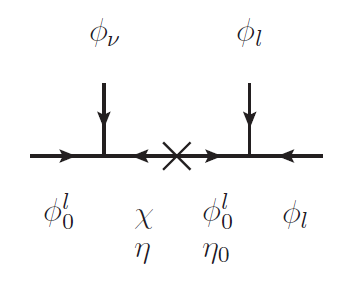}
\caption{Topologies for $\left\{ \phi_0^l \phi_l\phi_l \phi_\nu \right\}$. \label{fig:top_12}}
\end{figure}

While the minimal completion is viable, to some extent the relatively small values of $c_s/c_a$ that are allowed serve as motivation to check how the next-to-minimal completions fare.
We proceed similarly to section \ref{sec:AF}, but now explore in more detail the difficulties of perturbing the VEV alignment (the same type of issues also apply to the framework in section \ref{sec:AF}).
In each pair of messengers required for effective alignment terms, one of the fields has R-charge of $2$ i.e. often we add a new alignment field contributing new minimisation conditions.
Explicitly, consider the non-renormalisable terms $\xi' \left\{ \phi_0^l \phi_l \phi_\nu \right\}''$ and the different contractions $\left\{ \phi_0^l \phi_l\phi_l \phi_\nu \right\}$ that are invariant under $A_4 \times Z_4$. To enable them in a UV completion, we can introduce different messengers according to the topology of the respective diagram: 3 possibilities for $\xi' \left\{ \phi_0^l \phi_l \phi_\nu \right\}''$ (figure~\ref{fig:top_abc}) and 2 for $\left\{ \phi_0^l \phi_l\phi_l \phi_\nu \right\}$ (figure \ref{fig:top_12}).

Enabling at least one of the diagrams requires triplet messengers $\chi$, $\phi_0^l$ with respective $Z_4$ charges $-1$ and $-1$ or alternatively $\phi_l$, and $\chi'_0$ with $i,-i$; or $A_{4}$ singlet messengers $\eta$, $\eta_0$ with respective $Z_4$ charges $-1$ and $-1$ or alternatively $\eta'$, $\eta'_0$ with $i,-i$. The fields with subscript zero are the R-charge $2$ fields in each messenger pair. We have already identified some of the messengers as existing fields and depending on the $A_{4}$ singlet chosen for $\eta'_0$, the field $\xi'$ could serve as $\eta'$.

If one adds a R-charge 2 triplet such as $\chi'_0$, they add 3 new minimisation constraints to Eqs.(\ref{Malign_start})-(\ref{Malign_end}) and it can be verified that the new constraints eliminate the previous solution and lead to vanishing VEVs.
If one instead adds a R-charge 2 singlet such as $\eta_0$ together with $\eta$, it is possible to enable a single topology (see figure \ref{fig:top_12})
while preserving non-trivial VEVs. The new superpotential terms are $\left\{ \phi^0_l \phi_\nu \right\} \eta + \eta^0 (\eta + \left\{ \phi_l \phi_l \right\})$, and the new constraint added is
\begin{align}
\frac{\partial w}{\partial \eta_{0}} = \eta + \left\{ \phi_l \phi_l \right\}=0
\end{align}
where $\left\{ \phi_l \phi_l \right\} = \phi_{l1}^2 + 2 \phi_{l2}\phi_{l3}$. The new constraint is satisfied for $\langle \eta \rangle =0$ and $\langle \phi_l \rangle \propto (0,1,0)$. But as the singlet messengers only allow the topology where the $A_{4}$ contractions are $\left\{ \phi_l^0 \phi_\nu \right\} \left\{ \phi_l \phi_l \right\}$ and $\left\{ \phi_l \phi_l \right\} =0$, at the effective level the model remains unmodified.

The other singlet choice is the pair with charges $i,-i$. The minimal choice is $\eta' \equiv \xi'$ as this only requires adding $\eta'^0$ as a $1''$ of $A_{4}$. The new terms are $\eta'_0 \xi' + \eta'_0 \left\{ \phi_l \phi_\nu\right\}'$,
relating the VEV of $\xi'$ and the VEVs of the triplet flavons, which can be accommodated (originally $\langle \xi' \rangle$ was undetermined). An effective term $\left\{ \phi_0^l \phi_l \phi_l \phi_\nu \right\}$ is enabled but the singlet messengers only allow the $A_{4}$ contraction $\left\{ \phi^0 \phi_l\right\}'' \left\{ \phi_l \phi_\nu \right\}'$, where $\langle \left\{ \phi_l \phi_\nu \right\}' \rangle \propto \langle \xi' \rangle $ due to $\frac{\partial w}{\partial \eta'_{0}}=0$. We conclude that this choice amounts only to a redefinition of $h_1$ in Eqs.(\ref{Malign_start})-(\ref{Malign_end}).
If we place the $i,-i$ singlet messenger pair in other $A_{4}$ singlet irreps., we would modify Eqs.(\ref{Malign_start})-(\ref{Malign_end}) into equations quite similar to Eqs.(\ref{eqn:MinimizationAFModStart})-(\ref{eqn:MinimizationAFModEnd}), and thus $\phi_l$ would be forced into a $(1,1,1)$ VEV.
Instead, adding only $\eta'^0$ as a $1$ or $1''$ of $A_{4}$ without the corresponding $\eta'$ does not modify Eqs.(\ref{Malign_start})-(\ref{Malign_end}) while enforcing $\frac{\partial w}{\partial \eta'_{0}}=0$ leads either to $\left\{ \phi_l \phi_\nu  \right\}=0$ or $\left\{ \phi_l \phi_\nu \right\}'=0$ respectively for the $1$ and $1''$ choice for $\eta'^0$. This added constraint is inconsistent with the existing alignment equations and would lead to vanishing VEVs.

Finally, even the choice without an extra R-charge 2 field requires adding a triplet $\chi$ as $-1$ under $Z_4$. This enables two of the topologies in figures \ref{fig:top_abc} and \ref{fig:top_12}, through the terms:
$\left\{ \phi^0_l \chi  \right\} +  \left\{ \phi^0_l \chi  \right\}  \xi + \left\{ \phi^0_l \chi  \phi_\nu \right\}$. They perturb the existing equations without adding new ones. If $\langle \chi \rangle =0$ the contributions arise from the enabled contractions of the non-renormalisable terms $\left\{ \phi^0_l \phi_\nu \phi_l \right\}''  \xi' + \left\{ \phi^0_l \phi_\nu \phi_l \phi_l \right\}$. The alignment Eqs.(\ref{Malign_start})-(\ref{Malign_end}) get modified with extra terms that force $\langle \phi_l \rangle$ to be aligned in the same direction as $\langle \phi_\nu \rangle$.

With respect to modifying the alignment, even though \cite{Altarelli:2009kr} presents the NLO VEVs as $\langle \phi_\nu \rangle = \varepsilon(1+dw,1+dw,1+dw)$, $\langle \phi_l \rangle = (dx,u+dy,dz)$, we conclude that the next-to-minimal completions do not allow those general deviations for $\langle \phi_\nu \rangle$.

We consider now the remaining possibilities that do not involve modifications to the alignment. Trying to extend the Majorana or Dirac sector in this framework is not interesting: the non-renormalisable terms $\nu^c \phi_\nu \phi_l \nu^c$ and $l h_u \phi_l \nu^c$ are not invariant under $Z_4$, rather the terms would require $4$ extra insertions of the non-trivial $Z_4$ charged flavons ($\phi_l$ or $\xi'$). The additional messengers required to enable $4$ extra insertions push the associated completion beyond the next-to-minimal constructions we consider here. Instead, when more insertions of $\phi_\nu$ are added, one notes it only amounts to the seesaw mechanism of the minimal completion, as $\nu^c$ is the only messenger involved.

A final approach is to consider, analogously to section \ref{sec:AF}, an additional $A_{4}$ non-trivial singlet entering the renormalisable neutrino superpotential. In this case we add a $1''$ field, $\xi''$. It is easy to see that $\xi''$ should transform trivially under $Z_4$. At the effective level this field would produce a new NLO term for every term that has two or more $A_{4}$ triplets, but in the UV completion it appears in the following terms only
\begin{align}
\left\{ \nu^c \nu^c \right\}' \xi'' + \left\{ \phi^0_\nu \phi_\nu \right\}' \xi'' + \left\{ \chi_\tau^c \chi_\tau \right\}' \xi'' \,.
\label{eqn:xipp}
\end{align}
The first term is a new Majorana term that produces a similar effect to what was discussed in the end of section \ref{sec:AF}.
Similarly to what occurs with the alignment Eqs.(\ref{eqn:MinimizationAFModStart})-(\ref{eqn:MinimizationAFModEnd}), the alignment term $\left\{ \phi^0_\nu \phi_\nu \right\}' \xi''$ does not perturb the $(1,1,1)$ VEV for $\phi_\nu$.
The last term is a new contribution to the messenger mass of the $\chi_\tau^c$, $\chi_\tau$ pair.

\begin{figure}[t]
  \begin{center}
    \includegraphics[width=.8\textwidth]{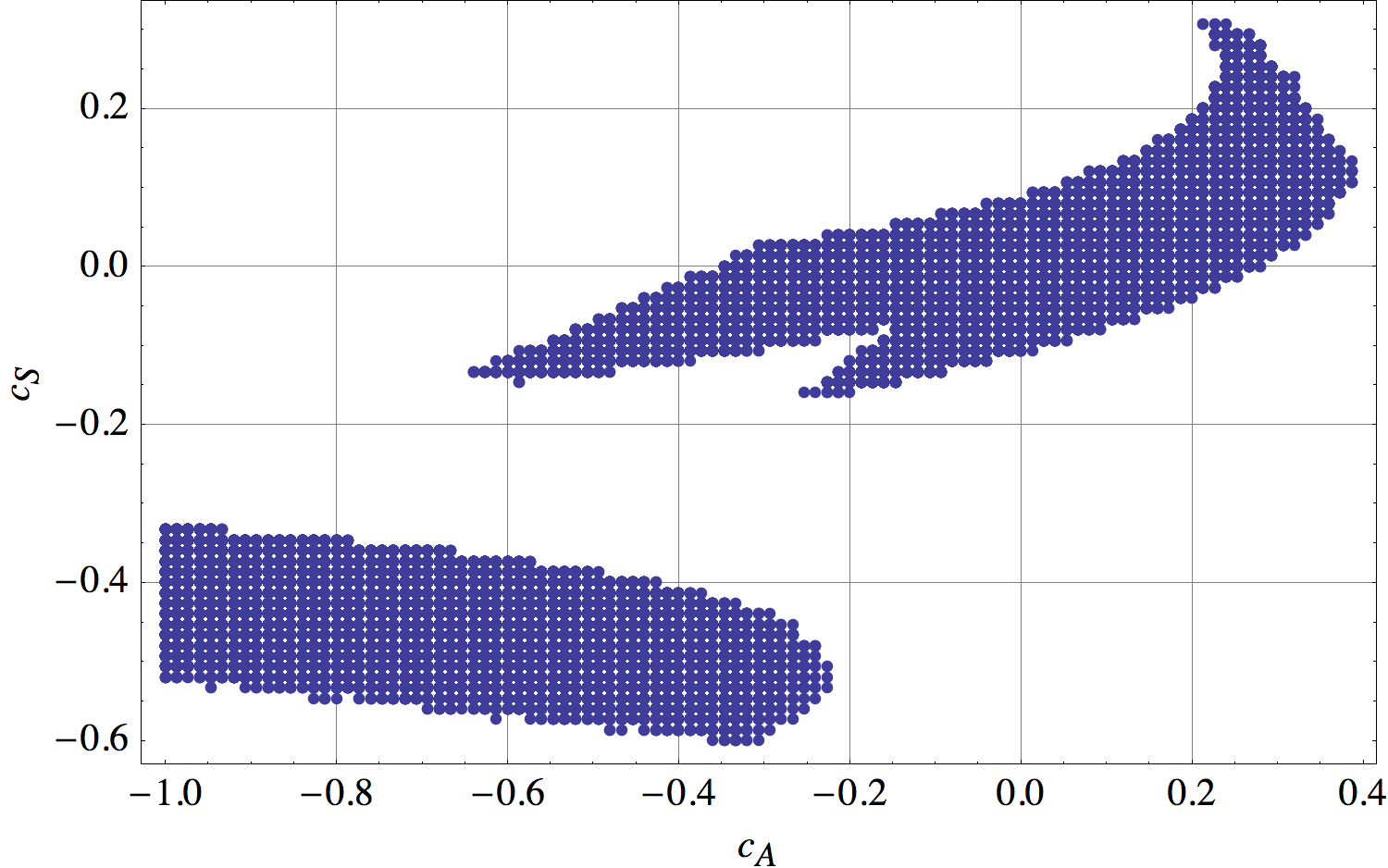}
  \end{center}
  \caption{Range of $c_{A}$ and $c_{S}$ allowed by present mixing angle data for the modified AM model ($\varepsilon' = \varepsilon''= 1$).}
  \label{fig:AMMod_CAvsCS}
\end{figure}

The phenomenological implications of Eq.(\ref{eqn:xipp}) appear through the combined effect of the modified messenger mass with the effect of the $\left\{ \nu^c \nu^c \right\}' \xi''$ term. Eq.(\ref{eqn:AM_mcl}) becomes
\begin{align}
  m_{l} &=  \begin{pmatrix}
    m_e & (-c_s+c_a) \varepsilon' m_\mu & (-c_s-c_a+c_{\chi}) \varepsilon'  m_\tau \\
    (-c_s+c_a+c_{\chi}) \varepsilon' m_e & m_\mu & 2 c_s \varepsilon' m_\tau \\
    \varepsilon' (-c_s-c_a) m_e & (2 c_s+c_{\chi}) \varepsilon' m_\mu & m_\tau
  \end{pmatrix},
  \label{eqn:AMModMl}
\end{align}
where $c_{\chi}$ denotes the coupling corresponding to the last term of Eq.(\ref{eqn:xipp}), while in the neutrino sector this adds at term with the structure of Eq.(\ref{eqn:MPP}) to the Majorana mass. 
As expected, the new flavon field relaxes the hierarchy in the parameters $c_{A}$ and $c_{S}$ which was necessary in the minimal model. This is illustrated in figure~\ref{fig:AMMod_CAvsCS}, which presents the allowed range for these parameters, which can now be of the same order of magnitude (c.f. with figure~\ref{fig:AM_CAvsCS}).

\section{Summary}

In the $A_{4} \times Z_3 \times U\left( 1 \right)_{\textrm{FN}}$ framework, the original minimal model is ruled out by large $\theta_{13}$ and the next-to-minimal completions do not fare much better. The most attractive alternative is the minimal UV complete model with a flavon transforming as a non-trivial $A_{4}$ singlet.
In the $A_{4} \times Z_4$ framework, the original minimal completion is viable but its parameters are strongly constrained by observations. We analysed in detail the different types of next-to-minimal constructions (e.g. modifications to the alignment). Again the most appealing alternative is having a non-trivial $A_{4}$ singlet flavon: this model is still constrained by observations but the constraints are relaxed.
As was already the case in \cite{Varzielas:2010mp}, one of the main conclusions is that the minimal and next-to-minimal UV complete models can be much more predictive than the respective non-renormalisable effective models, and indeed this enabled us to rule out several possibilities.

\acknowledgments

IdMV was supported by DFG grant PA 803/6-1 and partially supported through the project PTDC/FIS/098188/2008. DP was supported by DFG grant PA 803/5-1.

\bibliography{v2}
\bibliographystyle{apsrevM}

\end{document}